\documentclass[aps,pre,twocolumn,showpacs,floatfix,preprintnumbers,amsmath,amssymb,superscriptaddress,reprint]{revtex4-2}
\usepackage{amsmath}
\usepackage{amssymb}
\usepackage{graphicx}
\usepackage{epstopdf}
\usepackage{epsfig}
\usepackage{dcolumn}
\usepackage{bm}
\usepackage{url}
\PassOptionsToPackage{hyphens}{url}
\usepackage{hyperref}
\usepackage{natbib}
\usepackage{float}
\usepackage[caption = false]{subfig}

\begin{document}
\sloppy
\title{Inferring long memory using extreme events}
\author{Dayal Singh}
\email{dayal.singh@students.iiserpune.ac.in}
\author{M. S. Santhanam}
\email{santh@iiserpune.ac.in}
\affiliation{Indian Institute of Science Education and Research,\\
Dr. Homi Bhabha Road, Pune 411008, India.}

\date{\today}

\begin{abstract}
Many natural and physical processes display long memory and extreme events. In these systems, the measured time series is invariably contaminated by noise. As the extreme events display large deviation from the mean behaviour, the noise does not affect the extreme events as much as it affects the typical values. Since the extreme events also carry the information about correlations in the full time series, they can be used to infer the correlation properties of the latter. In this work, from a given time series, we construct three modified time series using only the extreme events. It is shown that the correlations in the original time series and in the modified time series, as measured by the exponent obtained from detrended fluctuation analysis technique, are related to each other. Hence, the correlation exponents for a long memory time series can be inferred from its extreme events alone. This approach is demonstrated for several empirical time series.
\end{abstract}

\maketitle

\section{\label{sec:level1} Introduction \protect}

Any event whose magnitude strays far from its typical values can be designated as an extreme event. Such extreme events have significant impact in both nature and society \citep{Albeverio2006, extreme-natural}. The consequences of naturally occurring extreme events such as the floods, droughts, cyclones and earthquakes are often disastrous. Extremely large solar flares such as the Halloween storms of 2003 \citep{nasa} can potentially damage the communication satellites, power distribution and communication networks and might require
re-routing of aircrafts \citep{Baker2008}. Due to our reliance on technology in day-to-day life, we encounter comparatively less disruptive extreme events ranging from mobile network congestion to traffic jams. In economy, market crashes have impacted the entire international financial system \cite{sornette}. Due to disproportionately large social and economic costs, it is essential to understand the extreme event properties and their early warning signals.

It is by now well understood that irrespective of the physical origins, extreme events display certain generic statistical properties. One such possibility arises in a large number of natural, socio-economic and technological systems display long memory property. This implies that the autocorrelation function decays sufficiently slowly in a power-law form, $\langle x(t) x(t+\tau) \rangle \sim \tau^{-\gamma}$ where $0< \gamma < 1$ is the autocorrelation exponent. Effectively, this points to absence of typical time-scales and also enchanced probability for temporal clustering of extreme events when compared to uncorrelated systems \cite{Santhanam2008,long1,long2,long3,long4}. Many systems which display extreme events, e.g., stock markets, atmospheric temperature, rainfall, earthquakes, physiological variables such as heart beat, electro encephalograph data, road traffic are long range correlated and fall in this class \cite{lmem, lmem1}. Primarily motivated by these examples, considerable research effort had been invested in studying the distribution of time interval $R$ between successive occurrences of extreme events called recurrence time distributions $P(R)$. For a time series with auto-correlation exponent $\gamma$, the approximate form for $P(R)$ has a power-law decay for short recurrence intervals, and a stretched exponential decay for long recurrence intervals \cite{Santhanam2008} and the characteristic exponents in these regimes are a function of $\gamma$. Thus, $\gamma$ carries information about the entire time series as well as about its extreme events.

Often, such characterisation through $\gamma$ becomes ambiguous when the long range correlated time series is contaminated by noise and/or missing data. For sufficiently strong noise contamination, a time series can lose its long range character and even become uncorrelated. Similarly, missing data also leads to uncorrelatedness. Generally, the extreme events strongly deviate from typical values and consequently are far less affected than the regular non-extreme events by the noise level in the measurements. Randomly missing data affects non-extreme events more than the extreme events since typically the non-extremes outnumber the extremes. All these arguments imply that it might be possible to study the statistical features of the extreme events alone and infer information about long-range correlations in a time series. Effectively, the information contained in the, possibly noisy, non-extreme values of the time series can be disregarded. Hence, the main premise of this article is to characterise a long range correlated time series by using only the extreme events. Further, apart from its application to noisy time series, in the context of the current interest in big data, this can be thought of as a method to estimate correlation exponents of very long time series using only a small fraction of its data that display extreme events.

In this work, we use detrended fluctuation analysis (DFA) to quantify long memory and the results are presented in terms of DFA exponent defined in Eq. \ref{fluctuation_relation}. This method has been extensively studied earlier, largely through simulations, and is useful even in the presence of non-stationarities \cite{chen2002,holl2016} and trends \cite{Hu2001} in the time series. Recently, detailed theoretical studies of DFA \cite{theory1, theory2, theory3} have shown that detrending is implicit if fluctuation function is to be an unbiased estimator \cite{holl2019}. Probabilistic approaches have been adopted to obtain expected values of fluctuation function for Gaussian processes \cite{sikora2020}. This paper could also be thought of as response of the long memory series to specific kind of data loss, and is relevant in the context of research interest in how DFA fluctuation functions behave under similar conditions of data loss \cite{Qianli2010}. In this work, we pick out extreme events in a time series for special treatment for their long memory properties. This has some broad parallels in earlier work in which sign and magnitude of the fluctuations have been singled out for special treatment as far as their long range properties are concerned \cite{manuel2016}. It must also be mentioned that DFA technique can be applied to classical and quantum systems at criticality \cite{landa2011}, and further it is equivalent to performing $\Delta_3$ statistics widely used in random matrix theory \cite{Santhanam2006}.

\begin{figure}[t]
\includegraphics*[width = 0.85\linewidth]{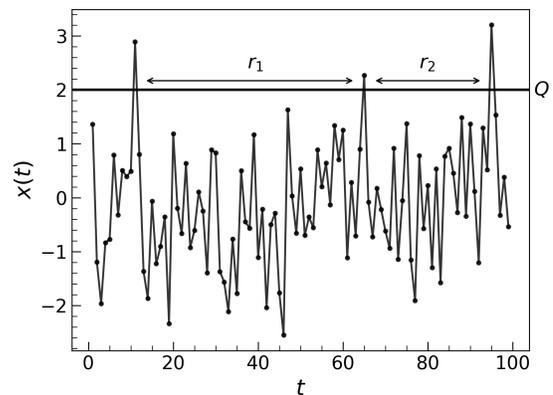}
\caption{A schematic time series $x(t)$. For a threshold of $Q = 2$, indicated by a horizontal line, three extreme events can be identified. Two return intervals, $r_1$ and $r_2$, are shown.}
\label{fig:ts}
\end{figure}

Let $x(t)$ be a long range correlated time series and $Q$ be the threshold which defines extreme events. Then, extreme events are those for which $x(t) > Q$. A schematic  of designating events as extreme events is shown in Fig. \ref{fig:ts}, in which the solid horizontal line defines the threshold for extreme events. This figure also shows recurrence intervals, the time interval between two successive extreme events. Suppose we consider Gaussian distributed timeseries $x(t)$ with long memory exponent $\gamma=0.2$, then upon adding white noise of sufficient strength, the series tends to become uncorrelated. As we show in this work, by isolating only the extreme events from $x(t)$, we can still infer about the long memory exponent of $x(t)$. Rest of the article is structured as follows : In Section \ref{sec:methods}, various methods and measured data used in this work is reviewed. In particular, we review detrended fluctuation analysis and the Fourier filtering method to generate synthetic long memory data. In Sections \ref{sec:extreme_events} to \ref{sec:comp_extreme_events}, we introduce three different methods of applying extreme events to infer about long memory exponent, and a regression based method is used to estimate the exponent in section \ref{sec:infer_dfa}. Finally, conclusions are presented in Section \ref{sec:conclusion}.

\section{\label{sec:methods} Methods and Data}

\subsection{Detrended Fluctuation Analysis}
\label{sec:dfa_int}
Detrended Fluctuation Analysis is a widely employed technique to quantify correlations in non-stationary time series data \citep{Peng1995, Bashan2008}. Several variants of this technique are also studied in the literature, see Ref. \cite{govindan2020} for one such variant based on orthogonal polynomials. We briefly review the basic technique here and the details are available in Ref. \cite{Peng1995}. 
Let $x(t'), t'=1,2,3,\dots,N$ denote a time series of length $N$ with mean $\mu$ and variance $\sigma^2$. The integrated version of the time series is given by
\begin{equation}
y(t) = \sum_{t' = 0}^t \: \left( x(t') - \mu \right).
\end{equation}
Now, $y(t)$ is partitioned into boxes of size $n$, where typically $n \le N/4$. Within each box, a polynomial $y_l(t)$ of order-$l$ is fitted to $y(t)$. In practice, usually order-$1$ is used. The time series is locally detrended by subtracting $y_i(t)$ from the integrated time series. The root-mean-square fluctuation function as a function of size of box $n$ is given by
\begin{equation}
F(n) = \sqrt{\frac{1}{N} \sum_{t = 0}^{n} \: \left( y(t) - y_l(t) \right)^2}.
\end{equation}
This process is repeated by varying the box size $n$. For correlated time series, the fluctuation function $F(n)$ generally scales as
\begin{equation}
F(n) \sim n^\alpha,
\label{fluctuation_relation}
\end{equation}
where $\alpha$ is the DFA exponent and indicates the degree of correlation. If $\alpha = 0.5$, then the series is uncorrelated. If $\alpha > 0.5$ then it is positively correlated (persistent) and if $\alpha < 0.5$ then it is anti-correlated (anti-persistent). The DFA method gives reliable result if $\alpha \in [0, 1]$ \citep{lmem,lmem1}, though it is also known to work for $\alpha > 1$ \citep{Extremera2016}. For anti-correlated series, the DFA method overestimates the exponent for small box sizes $n$ \citep{Hu2001}. In this case, $\alpha$ can be reliably estimated by first integrating the anti-correlated series and then applying DFA to it. The local trend is then removed by fitting a second-order polynomial as it is integrated twice in this process. The true exponent can be calculated from the estimated exponent $\alpha'$ using the relation $\alpha' = \alpha + 1$ \cite{Hu2001}. In Appendix \ref{Appendix_calibration}, it is shown that this method (we call it DFA-int) can estimate exponent when $\alpha < 0$. For stationary time series, $\alpha$ is related to the auto-correlation exponent $\gamma$, and power spectral exponent $\beta$ through the relations

\begin{eqnarray}
\label{eqn:alpha_beta}
\alpha = \frac{1+\beta}{2},\\
\label{eqn:alpha_gamma}
\alpha = 1-\frac{\gamma}{2}.
\end{eqnarray}
These relations \cite{Holl2015} are valid for $0 < \alpha < 1$ but it is worth mentioning that Eq. \ref{eqn:alpha_beta} also works for $\alpha > 1$ \citep{Extremera2016}. In Appendix \ref{Appendix_calibration}, this relation is extended to cover the range $-1 < \alpha < 0$ using the DFA-int method.

We use Fourier filtering method \citep{Makse1996, Rangarajan2000} to generate synthetic time series data $x(t)$ with desired DFA exponent $\alpha_{in}$.  Time series of length $2^{18}$ was generated and all the estimates for various exponents are averaged over 40 realizations. Using $x(t)$, the estimated DFA exponent is $\alpha_x$. In practice, $\alpha_{in} \approx \alpha_x$, and hence we use the latter in rest of this paper. Further, we employ surrogate data analysis \citep{Theiler1992} by randomly shuffling $x(t)$. If the shuffled series becomes uncorrelated, then it implies the presence of non-trivial correlations in the data and is not a chance occurrence.

\begin{table}[t]
\begin{tabular}{|c|c|c|c|}
\toprule

Data set & Years & Length of data  & DFA exponent   \\
\hline
S\&P 500 index  & 1927-2020 & 23248 & 0.865 \\[1pt] 
ED stock data   & 1962-2020 & 14739 & 0.850 \\[1pt]
IBM stock data  & 1962-2020 & 14739 & 0.883 \\[1pt]
BK stock data   & 1973-2020 & 11910 & 0.956 \\[1pt]
Seismic data    & 1981-2002 & 91797 & 0.784 \\[1pt]
Prague temperature & 1775-2019 & 89484 & 0.670 \\
\hline
\end{tabular}
\caption{Description of long-range correlated, empirical data analyzed in this work.}
\label{table1}
\end{table}

\subsection{Applications to observed data}
The results presented in this paper are tested using observed data sets. For this purpose, time series from three different systems are considered for extreme events based analysis. They are, ({\it a}) absolute log-returns of daily closing index and stock data, namely, S\&P index data, equity data of ED, IBM and BK stocks \citep{finance}, ({\it b}) seismic records from Italian Seismicity Catalog CSI 1.1 \citep{seismic}, and ({\it c}) observed daily mean temperatures from Prague observatory \citep{Bunde2001, govindan2003}. More details about the data are given in Table \ref{table1}. 

For the stock market data, if $x_k$ represents the value of stock/index at time $k$, then 
the absolute log-returns is defined as
\begin{equation}
\rho_k =  \left| \log_{10} \left( \frac{x_{k+1}}{x_k} \right) \right|.
\end{equation}
The extreme events analysis was peformed on $\rho_k$. 
The Italian seismicity catalog, CSI 1.1, contains magnitudes of earthquakes in the Italian territory during 1981-2002. The data has $91797$ earthquake records, of which $N = 39665$ have a magnitude evaluation. The observed daily temperature records from Prague observatory is also analyzed. Let $T_i$ represents the measured temperature on $i$-th day in any year. The seasonal trends were removed and we analysed the data of temperature anomaly $\Delta T$ defined by
\begin{equation}
\Delta T_i = T_i - \overline{T}_i,
\end{equation}
where $\overline{T}_i$ is the average temperature for the $i$-th day taken over all years. The correlation properties of this series had been studied earlier \cite{govindan2003}. 

As seen from the last column of Table \ref{table1}, the DFA exponent values indicate that all the time series are long range correlated. In all the cases, the time series were standardised to zero mean and unit variance. To define extreme events in these measured data sets, a cut-off defined by $q=1$ was chosen (see Eq. \ref{eqn:threshold} below for further explanation).

\section{\label{sec:extreme_events} Extreme events : modified time series}

\begin{figure}[t]
\includegraphics[width=0.85\linewidth]{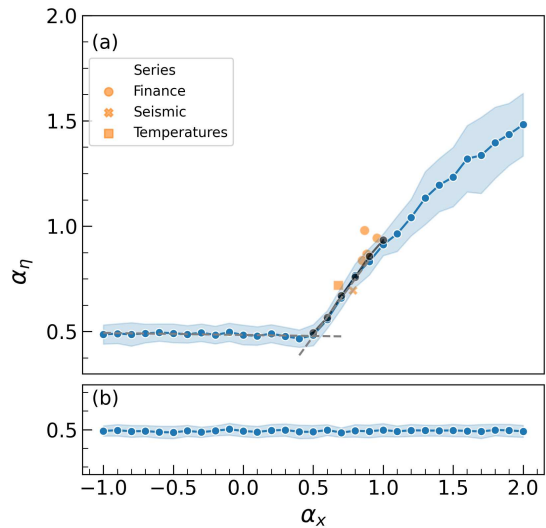}
\caption{\label{fig:dfa_ee} (a) Estimated DFA exponent $\alpha_\eta$ of extreme event series $\eta(t)$ as a function of $\alpha_x$ for synthetic data. The shaded region shows error bar around the mean trend. The two dashed lines are obtained by regression for the anti-correlated and the long-range correlated ($0.5 < \alpha_x < 1$)regimes with slopes, respectively, $m_1 = -0.009$ and $m_2 = 0.907$. The orange symbols, from the analysis of observed data, closely follow the mean trend. The black symbols are the DFA exponents for synthetic data with uniform noise added. (b) DFA exponent of randomly shuffled $\eta(t)$.}
\end{figure}


In this section, we analyse a modified time series in which only the extreme events are retained and all others are set to zero. A correlated time series $x(t)$ with DFA exponent $\alpha_x$ is considered. It is assumed that $x(t)$ has a well defined mean $\mu$ and standard deviation $\sigma$. An event at any instant of time $t$ will be designated as an extreme event if $x(t) > x_{th}$, where the threshold is 
\begin{equation}
x_{th} = \mu + q ~ \sigma
\label{eqn:threshold}
\end{equation}
and $q \ge 0$. From the given original time series $x(t)$, a modified time series $\eta(t)$ is constructed as follows :
\begin{equation}
    \eta(t) = 
\begin{cases}
    x(t), & \text{if } x(t) > x_{th} \\
    0,              & \text{otherwise}.
\end{cases}
\label{eqn:ee}
\end{equation}
Throughout this paper, we choose $q = 1$ as a threshold for extreme events. Further, we address the question -- given long range correlated time series $x(t)$ with DFA exponent $\alpha_x$, what is the DFA exponent $\alpha_\eta$ of the modified time series $\eta(t)$ ? Thus, we explore the self-affinity of $\eta(t)$ using DFA technique for $\alpha_{x} \in [-1, 2]$. The result relating $\alpha_x$ and $\alpha_\eta$, for which evidence will be presented below, can be stated as follows :
\begin{align}
    \alpha_\eta & \approx 0.5, \;\;\;\;\; -1 \le \alpha_x \le 0.5, \nonumber \\
                & < \alpha_x,  \;\;\;\;\;\;\; 0.5 < \alpha_x \le 2.
\label{eqn:res1}
\end{align}
This result is presented in Fig. \ref{fig:dfa_ee}(a). In this figure, the original long range correlated time series $x(t)$ of length $2^{18}$ is synthetically generated using Fourier filtering technique. By putting $q=1$, extreme events are identified and a modified time series $\eta(t)$ is generated. The DFA technique is applied to $\eta(t)$. Figure \ref{fig:dfa_ee}(a) shows $\alpha_\eta$ plotted as a function of $\alpha_x$. As can be inferred from this figure, the modified series $\eta(t)$ is found to be uncorrelated in the anti-correlated regime of $x(t)$, {\it i.e.,} $\alpha_x < 0.5$, whereas the correlations increase monotonically in the positively correlated regime, $\alpha_x > 0.5$, with an approximate slope of about 0.907. This systematics imply that $\alpha_\eta$ can be used to estimate the value of $\alpha_x$ in the persistence regime with $\alpha_x \in [0.5,1]$. However, if $\alpha_x > 1$ (strongly correlated regime), $\alpha_\eta$ is not reliable due to large variance primarily due to the finite size of the time series. 
Figure \ref{fig:dfa_ee}(a) also displays a similar analysis performed on observed data sets listed in Table \ref{table1}. It shows that $\alpha_\eta$, for observed data sets, closely follows the trend obtained from synthetic long range correlated data. This further confirms the systematic relation between $\alpha_{\eta}$ and $\alpha_x$. 

To physically understand these results, we realise that Eq. \ref{eqn:ee} is essentially a process of data loss. For an anti-correlated series, even moderate data loss of about $\sim 10\%$ is known to lead to significant change in the DFA exponent \citep{Qianli2010}. In the simulations presented in Fig. \ref{fig:dfa_ee}(a) nearly $\sim 84\%$ (for $q = 1$) of the series is removed and hence we expect the short range correlations in the rest of the series to be destroyed. This results in an uncorrelated time series. However, remarkably, the long-range correlated series is relatively more robust against data loss of non-extreme events. This is due to the fact that persistence ensures that extreme events are bunched together and the construction in Eq. \ref{eqn:ee} does not destroy these extreme event values, and as a result their correlations are preserved with slight modifications. 
The effect of noisy data is also shown in Fig. \ref{fig:dfa_ee}. The data $\eta(t)$ is deliberately contaminated with uniformly distributed noise $\xi \in [-0.5,0.5]$. The black symbols are the DFA exponents for noisy $\eta(t)$. As this reveals, the additive noise in the data does not significantly affect the value of $\alpha_{\eta}$ since extreme events are less affected by noise than normal events. Finally, Figure \ref{fig:dfa_ee}(b) shows the DFA exponent obtained after shuffling the time series. As expected, it shows that $\alpha_\eta \approx 0.5$ confirming that shuffling process had destroyed all correlations, and that result in Fig. \ref{fig:dfa_ee}(a) is not a chance occurrence.

\section{Correlations in return intervals series}
\label{ret_int_fluct}

The time ordered extreme events, exploited in section \ref{sec:extreme_events}, is one useful piece of information about the time series. Another useful component of the time series is the return intervals between successive occurrences of extreme events. This had been extensively studied as recurrence statistics, and it is fairly well understood that the return interval distribution is parameterised by the auto-correlation exponent $\gamma$ of a long range correlated time series \cite{Santhanam2008}. However, as shown in Appendix \ref{Appendix_cross_correlations}, the return intervals are not correlated with the extreme event values as measured by the linear Pearson correlation. These two components of a long range correlated time series, the extreme event values and the recurrence statistics, are independent of one another.

\begin{figure}[t]
   \includegraphics[width=0.85\linewidth]{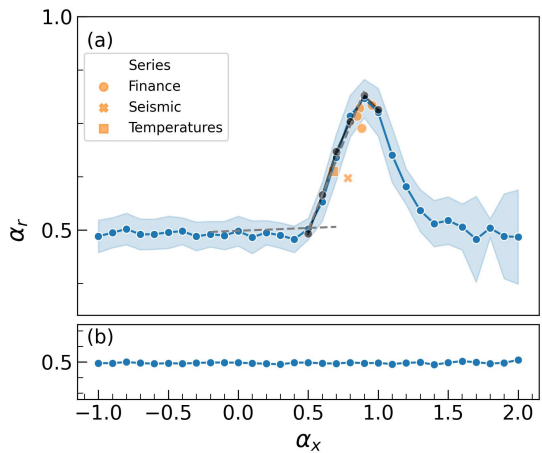}
\caption{\label{fig:dfa_ri} (a) Estimated DFA exponent of return interval series, $r(t)$ plotted against $\alpha_x$. The shaded region shows the error bar around the mean trend. The two dashed lines obtained by regression for the anti-correlated and long-range correlated regimes have slopes $m_1 = 0.013$ and $m_2 = 0.811$ respectively. The orange symbols are from observed data sets in Table \ref{table1}. The black symbols are the DFA exponents for synthetic data with uniform noise added. (b) DFA exponent of randomly shuffled $r(t)$.}
\end{figure}

For any $x(t), t=1,2,3, \dots,N$, with respect to extreme event threshold $x_{th}(q)$, we consider two successive occurrences of extreme events at times $t=t_{m}$ and $t_{m+1}$. Then, the return interval series is defined as
\begin{equation}
r(m) = t_{m+1} - t_m, \;\;\;\;\; m=1,2,3,\dots, 
\label{eqn:ri}
\end{equation}
and its DFA exponent is denoted by $\alpha_r$. Figure \ref{fig:dfa_ri} displays simulation results for how $\alpha_r$ varies as a function of $\alpha_x$. It reveals that the return intervals are uncorrelated in the anti-correlated regime of $\alpha_x$. This is to be expected since the extreme events are uncorrelated in this regime, and not surprisingly the return intervals are uncorrelated. In contrast, in the regime of $0.5 <\alpha_x < 1$, $\alpha_r$, increases approximately linearly (till about $\alpha_x = 0.9$) at a rate slower than for $\alpha_{\eta}$. In general, we infer that,
\begin{equation}
\label{eqn:ri_x}
\alpha_r < \alpha_\eta <\alpha_x.
\end{equation}
Upon further increasing beyond $\alpha_x>1$, the correlations begin to monotonically decrease, and the return intervals become almost uncorrelated. This happens because as $\alpha_x$ increases, correlations get stronger, and more and more extreme events are consecutive in time with return interval $r=1$. Then, due to consecutive extreme events, $r(m)=1$ for long times and this suppresses fluctuations leading to a decrease in $\alpha_r$. In fact, this effect appears to take over even as $\alpha_x \to 0.9$. For high values of $\alpha_x>1.5$, the return interval correlations fluctuate about $\alpha_r = 0.5$, as most of the series consists of consecutive ones. The relatively high variance in this regime is due to the finite length of the series, leading to high variability in the number of consecutive extreme events present. As $m \to \infty$, the return interval series will be slightly correlated in the regime $\alpha_x > 1.5$. Figure \ref{fig:dfa_ri}(a) also displays the values of $\alpha_r$ computed for the observed data sets listed in Table \ref{table1}. A good agreement is observed with the trend shown by the synthetic data (blue symbols). The systematic relationship between $\alpha_r$ and $\alpha_x$ is also exhibited by the observed time series. We might also point out that the additive noise (uniform noise added to $r(t)$) in the data does not significantly affect the value of $\alpha_{r}$, as seen by the black symbols in Fig. \ref{fig:dfa_ri}(a). If the return interval series $r(t)$ is randomly shuffled, the DFA exponent is approximately 0.5 (Fig. \ref{fig:dfa_ri}(b)) pointing to the existence of non-trivial correlations in $r(t)$.

\begin{figure}[t]
   \includegraphics[width=0.85\linewidth]{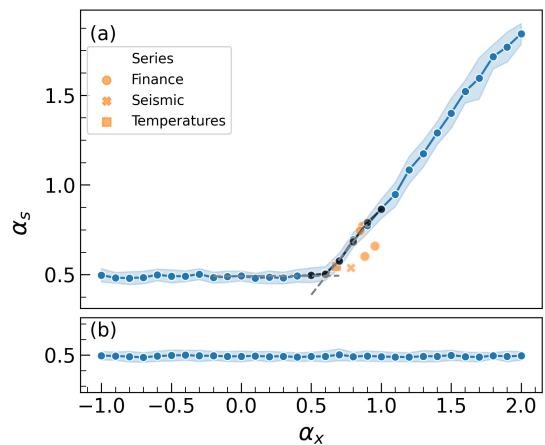}
\caption{\label{fig:dfa_ee_compressed} (a) Estimated DFA exponent of compressed extreme event series, $s(t)$, plotted against $\alpha_x$. The shaded region represents the error bar. The two dashed lines obtained by regression for the regimes, $\alpha_x < 0.5$ and $0.5 < \alpha_x < 1$, have slopes $m_1 = 0.003$, $m_2 = 0.983$  respectively. The orange symbols represent $\alpha_s$ for the observed data listed in Table \ref{table1}. The black symbols are the DFA exponents for synthetic data with uniform noise added. (b) DFA exponent of randomly shuffled time series $s(t)$.}
\end{figure}

\section{Compressed extreme events}
\label{sec:comp_extreme_events}
Starting from $x(t)$, we construct a time series $s(m)$ in which only the extreme events are retained by removing all instances of non-extreme events. We call this compressed extreme event series and is defined as
\begin{equation}
s(m) = \{\: x(t_m) \: | \:x(t=t_m) > x_{th} \: \}, \;\;\;\; m=1,2,3 \dots N_e,
\label{eqn:compressed_ee}
\end{equation}
where $N_e$ is the number of extreme events in the time series $x(t)$ of length $N$. In most cases, $N \ge N_e$. Note that in contrast to Eq. \ref{eqn:ee}, the non-extremes in Eq. \ref{eqn:compressed_ee} are entirely removed instead of setting them to zeros. The DFA exponent associated with the time series $s(m)$ is denoted by $\alpha_s$.
The variation of $\alpha_s$ as a function of $\alpha_x$, is shown in Fig. \ref{fig:dfa_ee_compressed}. Evidently, the compressed extreme event series $s(m)$ is  uncorrelated in the anti-correlated regime, $\alpha_x < 0.5$. This is the result of extreme event series $x(t)$ being nearly uncorrelated.

In the correlated regime, $\alpha_x > 0.5$, the observations shown in Fig. \ref{fig:dfa_ee_compressed} reveal that 
$\alpha_s < \alpha_x$. This can be understood as follows. For the series $x(t)$, the fluctuation function is denoted as $F(L) \sim L^{\alpha_x}$. For a time series $x(t)$ of length $L$, the corresponding 
length of the compressed time series
is $L' = L/\langle r \rangle$, where $\langle r \rangle$ represents the mean return interval for extreme events, which depends on the $x_{th}$ used to identify extreme events.
Then, using the equation for $F(L)$, the fluctuation function for compressed extreme events can be written as
\begin{align}
F(L') & \sim \left( \frac{L'}{ \langle r \rangle } \right)^{\alpha_x} \nonumber \\
      & \sim L'^{\; \alpha_x (1-\beta)}, 
\label{Fcompee}
\end{align}
where the DFA exponent of the compressed series $s(m)$ can be identified as $\alpha_s = \alpha_x (1-\beta)$ and
box-size dependent $\beta$ is,
\begin{equation} 
  \beta = \log_{L'} \langle r \rangle = \frac{\log \langle r \rangle}{\log L'}.
\label{beta}
\end{equation}
In the limit of large box size $L'$, {\it i.e.} , $L' >> 1$, we have $\beta << 1$. Hence, we can 
infer that $\alpha_x > \alpha_s$, as observed in the simulation results displayed in Fig. \ref{fig:dfa_ee_compressed}.
The dependence on box size $L'$ is logarithmic and hence sufficiently weak that for finite sample sizes
the DFA exponent of the compressed extreme event series does not change appreciably with length of the time series. As predicted by Eq. \ref{beta}, simulation results (not shown here) verify that $\alpha_s \approx \alpha_x$ as $L' \to \infty$.

Based on Eqns \ref{Fcompee}-\ref{beta} and Fig. \ref{fig:dfa_ee_compressed}, the relation between $\alpha_x$ and $\alpha_s$ can be surmised as,
\begin{align}
    \alpha_s & \approx 0.5, \;\;\;\;\; -1 \le \alpha_x \le 0.5, \nonumber \\
                & < \alpha_x,  \;\;\;\;\;\;\; 0.5 < \alpha_x \le 2.
\label{eqn:res1}
\end{align}
For $0.5 < \alpha_x < 1$, $\alpha_x$ and $\alpha_s$ bear a linear relation between them. However, $\alpha_x  > 1.5$, correlations are far stronger and most extreme events tend to be consecutive with $\langle r \rangle \approx 1$. Hence, $\beta \approx 0$ and $\alpha_x \approx \alpha_s$. Thus, for highly correlated time series, $\alpha_s$ provides a reasonable estimate of $\alpha_x$. Fig. \ref{fig:dfa_ee_compressed}(a) also shows $\alpha_s$ computed for the observed data sets in Table \ref{table1}. A reasonably good agreement is seen between the trends of synthetic data (blue symbols) and the observed data (orange symbols). If additive noise (uniform noise added to $s(t)$) is present in the data, even then it does not significantly affect the value of $\alpha_{s}$, as seen by the black symbols in Fig. \ref{fig:dfa_ee_compressed}(a). This is due to extreme events being less affected by noise than the non-extreme events. Finally, as shown in Fig. \ref{fig:dfa_ee_compressed}(b), if the compressed extreme event series $s(t)$ is randomly shuffled, its DFA exponent is nearly 0.5 showing that the compressed extreme events carry non-trivial correlations inherited from the original time series $x(t)$.

\begin{figure}[t]
   \includegraphics*[width=1.0\linewidth]{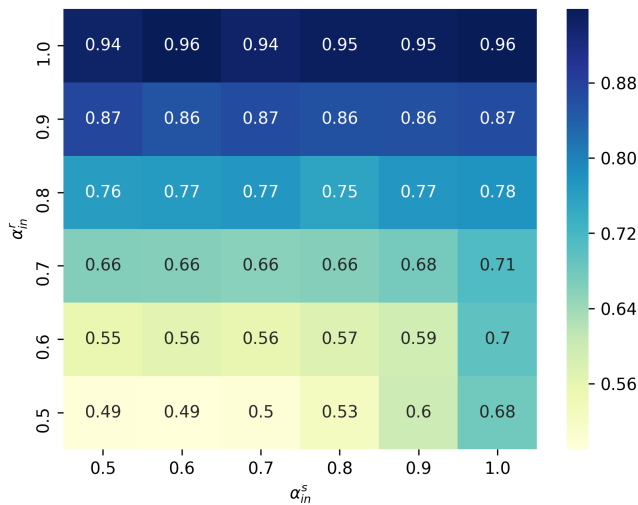}
\caption{The heat map shows the DFA exponent of the time series obtained by assembling two components together; the return intervals with DFA exponent $\alpha_{in}^r$, and the extreme events series with DFA exponent $\alpha_{in}^s$. Notice that the composite time series has DFA exponent close to that of its return intervals. The numbers in the grid give the actual DFA exponent, and the colormap encodes the same information.}
\label{fig:composite}
\end{figure}

Finally, it must be pointed out that return intervals series $r(t)$ and the compressed extreme event series $s(t)$ are uncorrelated processes, for most part. The linear correlation between them (not shown here) reveals that they are nearly uncorrelated. From any time series $x(t)$, the derived series $r(t)$ and $s(t)$ are created by considering extreme events defined with, say, $q=1$. This approach also provides a way of performing the reverse process -- assembling a time series with specified return interval correlations and specified extreme event correlations whose DFA exponents, respectively, are $\alpha_{in}^r$ and $\alpha_{in}^s$. If we assemble such a time series, then the DFA exponent obtained is displayed in Fig. \ref{fig:composite}. This figure reveals that for a time series $x(t)$ with DFA exponent $\alpha_x$, the correlations largely are contributed from the return intervals (with exponent $\alpha_{in}^r$) and is only weakly by the correlations in the extreme event values (with exponent $\alpha_{in}^s$).

\section{Inferring DFA exponent using extreme events}
\label{sec:infer_dfa}
The central premise of this paper is that the information about temporal correlations, or equivalently the DFA exponent, of any time series is also embedded in its extreme events. Hence, using only the values of extreme events, we can infer the DFA exponent of the original time series. In this section, we employ the systematic relation between the DFA exponents that depend only on the extreme events, namely, $\alpha_{\eta}, \alpha_r$ and $\alpha_s$ and that for the original time series $\alpha_x$, depicted in Figures \ref{fig:dfa_ee}-\ref{fig:dfa_ee_compressed}, to infer the value of $\alpha_x$. In principle, any of $\alpha_i, i=\eta, r, s$ can be used for this purpose because $\alpha_i$ is a monotonically increasing function of $\alpha_x$ in the regime when $0.5 < \alpha_x < 1$. In order to tightly constrain the estimated exponent $\widetilde{\alpha}_x$, we propose a simple regression procedure that uses all the available information as outlined below.

Let $x(t), t=1,2 \dots T$ represent a {\sl persistent} time series for which we need to estimate the DFA exponent under the condition that the sample is noisy and it can be safely assumed that the extreme events are less affected by noise than non-extreme events. First step would be to decide a suitable threshold, $x_{th} = \langle x \rangle + \sigma_x$, where $\langle x \rangle$ and $\sigma_x$ are, respectively, the mean and standard deviation of the time series. Then, based on the extreme events so identified, three different time series $\eta(t), r(t)$ and $s(t)$ are constructed and, respectively, their DFA exponents $\alpha_{\eta}^{*}, \alpha_r^{*}$ and $\alpha_{s}^{*}$ are computed. For $0.5 < \alpha_x < 1$, it is clear from Figs. \ref{fig:dfa_ee}-\ref{fig:dfa_ee_compressed} that the estimated exponents have a systematic dependence on $\alpha_x$ and hence a cost function $U$ that depends on $\alpha_x$ can be defined as,
\begin{equation}
U(\alpha_x) = \sqrt{(\alpha_{\eta}^{*} - \alpha_{\eta})^2 + (\alpha_{r}^{*} - \alpha_{r})^2 + (\alpha_{s}^{*} - \alpha_{s})^2}.
\label{eq:costfunction}
\end{equation}
Now, we set $dU/d\alpha_x = 0$, and obtain an estimate for DFA exponent to be 
\begin{equation}
\widetilde{\alpha}_x = g\left(\alpha_{\eta}^{*}, \alpha_r^{*}, \alpha_s^{*} \right).
\label{eq:estimate_alpha}
\end{equation}
where $g(.)$ must be obtained numerically. In Table \ref{table2}, the results of this procedure are tabulated for the measured time series listed in Table \ref{table1}. It is seen that, within the limitations of finite data length, a reasonable agreement is obtained between the actual DFA exponent and the estimated exponent. In principle, this procedure need not necessarily use the information of all the three estimated exponents $\alpha_{\eta}^{*}, \alpha_r^{*}$ and $\alpha_s^{*}$. However, using all these three exponents provides a tighter constraint for $\widetilde{\alpha}$. In the anti-persistent regime when $0 < \alpha_x < 0.5$, the DFA exponents computed from the derived time series $\eta(t), r(t)$ and $s(t)$ are $\alpha_{\eta} \approx \alpha_r \approx \alpha_s \approx 0.5$. Hence, the regression procedure would not estimate the correct value of DFA exponent, but would still yield the qualitative information that the original time series is anti-persistent in nature.

\begin{table}[t]
\begin{tabular}{|c|c|c|c|}
\toprule

Data set & DFA exponent & Estimated exponent \\
         &              & using extreme events \\
\hline
S\&P 500 index  & 0.865 & 0.972 \\[1pt] 
ED stock data   & 0.850 & 0.878 \\[1pt]
IBM stock data  & 0.884 & 0.799 \\[1pt]
BK stock data   & 0.956 & 0.881 \\[1pt]
Seismic data    & 0.784 & 0.675 \\[1pt]
Prague temperature & 0.670 & 0.679 \\
\hline
\end{tabular}
\caption{A comparison between "exact" DFA exponent $\alpha$ and the value estimated using extreme events
as described in Sec \ref{sec:infer_dfa} through Eqns. \ref{eq:costfunction} and \ref{eq:estimate_alpha}.}
\label{table2}
\end{table}

\section{\label{sec:conclusion}Summary}

Extreme events are the ones that deviate strongly from typical events. Generally, methods such as the detrended fluctuation analysis or its other variants are used to determine the correlation exponent of the long memory time series. In any typical time series, though the extreme events are fewer in number compared to typical events, the former have disproportionate influence in the time series. The measured time series could often be contaminated by noise and missing data points. Typically, the extreme events, due to their large magnitudes, are not as much affected by noise as the non-extreme events. Hence, in this work, we exploit this property by attempting to estimate the correlation exponent of a time series from its extreme events alone. we have systematically removed non-extremes in a correlated time series and the effect of removal on the correlation exponent has been studied. For this, given a time series $x(t)$, three different time series have been constructed that contain information about extreme events alone. We show that the DFA exponents for the original time series $x(t)$ and of the three derived time series (based on extreme events alone) collectively denoted by $x_{ee}(t)$ have a systematic relation. This result, in Figs. \ref{fig:dfa_ee}-\ref{fig:dfa_ee_compressed}, has been numerically obtained for extreme events defined with $q=1$ (see Eq. \ref{eqn:threshold}). However, our results indicate that qualitatively similar trends are observed for other values of $q$ as well. We use this relation to estimate the DFA exponent of the original time series, effectively using only the information about the extreme events in them. A regression based method is put forward to estimate the DFA exponent of $x(t)$. This technique has been demonstrated on several measured real time series data from physical systems, namely, stock markets, seismic and temperature data, all of which are known to be long range correlated and also display extreme events. The estimated exponents based on $x_{ee}(t)$ are in good agreement with the exponents obtained from the measured time series. The work presented is an exploration of the relationship between a time series and its extreme events treated as a derived time series. This raises several interesting questions that require a deeper investigation such as the role played by the extreme events thresholds in inferring DFA exponents, study of time series in which the extreme events magntiudes and the return intervals are correlated. Some of these will be explored and reported elsewhere.

\begin{figure}[t]
\includegraphics[width = 0.85\linewidth]{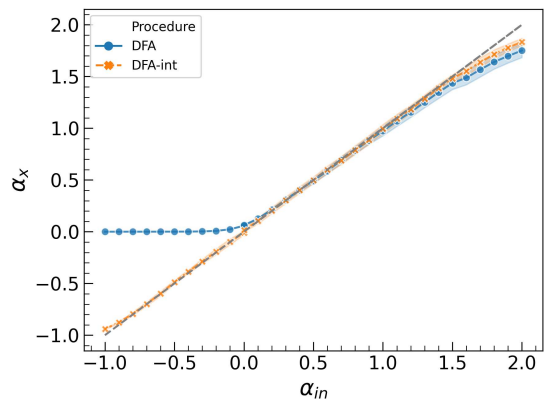}
\caption{\label{fig:dfa_synthetic} Estimated DFA exponent, $\alpha_x$, against the input DFA exponent $\alpha_{in}$ for the two methods. The shaded region shows the error bar. Deviation from grey line would imply that the method does not work well. The DFA-int gives reliable results in the range $[-0.7, 1.5]$ whereas the standard DFA method is reliable only in $[0.3, 1.0]$.}
\end{figure}

\begin{acknowledgments}
Dayal Singh would like to acknowledge the INSPIRE-SHE programme of Department of Science \& Technology, India. M. S. Santhanam acknowledges the MATRICS grant MTR/2019/001111 from SERB, Govt. of India.
\end{acknowledgments}

\appendix
\section{Benchmarking DFA-int in the extended range of correlations}
\label{Appendix_calibration}

In this, we study the scaling behaviour of correlated series in the extended range of the DFA exponent, i.e, $\alpha_x \in [-1, 2]$. This corresponds to the power spectral exponent $\beta \in [-3, 3]$. 

Correlated series of length $2^{18}$ was generated from a Gaussian distribution with zero mean and unit variance using the Fourier filtering method having DFA exponent $\alpha_x$. The input DFA exponent $\alpha_{in}$ was calculated using Eqn. \ref{eqn:alpha_beta}). The exponent $\alpha_x$ is estimated by two methods, the standard DFA and and DFA-int (as described in section \ref{sec:methods}), for the entire range and the trends are shown in Fig. \ref{fig:dfa_synthetic}. The results in this figure are averaged over 20 realizations.

As seen in Fig. \ref{fig:dfa_synthetic}, DFA-int gives reliable agreement with $\alpha_{in}$ in a wider range $[-0.7, 1.5]$, whereas the standard DFA method is reliable only in a shorter range $[0.3, 1]$. The standard DFA procedure outputs $\alpha_x = 0$ for all values of $\alpha_{in} < 0$ whereas DFA-int gives exactly the same exponent for almost all values of $\alpha_{in} < 0$. Now the procedure overestimates the DFA exponent near $\alpha_{in} \approx -1$. Both the procedures underestimate the DFA exponent for $\alpha_{in} > 1.5$, though DFA-int always provides a better estimate.

\begin{figure}[t]
\includegraphics[width = 0.85\linewidth]{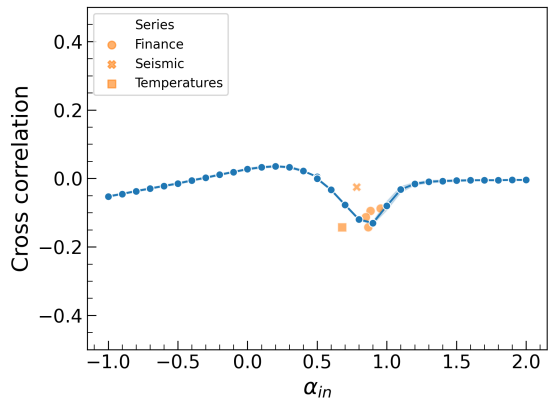}
\caption{\label{fig:cross} Numerically estimated linear Pearson correlation coefficient between return intervals and extreme event, as a function of the DFA exponent of $x(t)$. }
\end{figure}

\section{Correlations between return intervals and extreme event magnitudes}

\label{Appendix_cross_correlations}
Consider a long memory time series $x(t)$. The extreme events in this series are denoted by $y_i=x(t), i=1,2,3,\dots$, such that $x(t) > x_{th}$, where $x_{th}$ is the threshold identified in Eq. \ref{eqn:threshold}. Corresponding to every $y_i$, a return interval $r_i$ can be identified as the time elapsed since the last occurrence of extreme event. The cross-correlation between $y_i$ and $r_i$ is computed and the result is shown in Fig. \ref{fig:cross}. As seen in this figure, the return intervals and extreme events display mild anti-correlations for $0 < \alpha_x < 1$, and is nearly uncorrelated outside this regime of $\alpha_x$. Hence, to a first approximation, the extreme events and return intervals can be taken to be approximately uncorrelated.


\end{document}